\documentclass[12pt]{article}

\usepackage{graphics}
\usepackage[dvips]{graphicx}
\usepackage[left]{lineno}

\usepackage[dvipsnames]{xcolor}
\usepackage{ulem}

\definecolor{Boxbackground}{rgb}{0.91, 0.839, 1}       

\newif\ifTrackChange

\usepackage{amsmath} 
\usepackage{setspace} 
\usepackage{hyphenat}
\usepackage{geometry}
\usepackage{enumitem}

\usepackage{hyperref}
\hypersetup{
  colorlinks   = true,  %Colours links instead of ugly boxes
  urlcolor     = blue,  %Colour for external hyperlinks
  linkcolor    = Brown, %Colour of internal links
  citecolor    = Brown  %Colour of citations
}
\usepackage[framemethod=TikZ]{mdframed}%{framed}

\title{Architecture of Environmental Risk Modelling: for a faster and more robust response to natural disasters}

\usepackage{ifthen}
\ifthenelse{1=1}{
\usepackage{fancyhdr}
\pagestyle{fancy}
\fancyhf{} 

\fancyfoot[L]{\ifthenelse{\value{page}=1}{{\footnotesize \vspace{2mm}Cite as:\\
Rodriguez-Aseretto, D., Schaerer, C., de Rigo, D., 2014. \textbf{Architecture of Environmental Risk Modelling: for a faster and more robust response to natural disasters}. \textit{$3^{rd}$~Conference of Computational Interdisciplinary Sciences}, Asunci\'on, Paraguay
}}{}}
\fancyfoot[C]{\ifthenelse{\value{page}=1}{}{\thepage}}
\fancyfoot[R]{\ifthenelse{\value{page}=1}{\thepage}{}}
\fancyhead[R]{}
\fancyhead[L]{\ifthenelse{\value{page}=1}{}{\footnotesize Rodriguez-Aseretto, D., Schaerer, C., de Rigo, D., 2014. \textit{Architecture of Environmental Risk Modelling: for a faster and more robust response to natural disasters}. CCIS 2014}}
}

\begin{document}

\begin{center}

{\large{\bf Architecture of Environmental Risk Modelling: for a faster\\[1mm]

and more robust response to natural disasters}}
\vspace{1mm}
\bigskip

{\small Dario Rodriguez-Aseretto~$^{a,}$\footnote{E-mail Corresponding Author: dario.rodriguez@jrc.ec.europa.eu}, Christian Schaerer~$^{b}$, Daniele de Rigo~$^{a,c}$ 
}
\bigskip

{\small
$^a$ European Commission, Joint Research Centre, Institute for Environment and Sustainability, Ispra (VA), Italy.\\
$^b$ Polytechnic School, National University of Asuncion, San Lorenzo, Central, Paraguay.\\
$^c$ Politecnico di Milano, Dipartimento di Elettronica, Informazione e Bioingegneria, Milano, Italy.
}

\end{center}
\vspace{0mm}

\begin{abstract}
\noindent Demands on the disaster response capacity of the European Union are likely to increase, as the impacts of disasters continue to grow both in size and frequency.
This has resulted in intensive research on issues concerning spatially-explicit information and modelling and their multiple sources of uncertainty. Geospatial support is one of the forms of assistance frequently required by emergency response centres along with hazard forecast and event management assessment. 
Robust modelling of natural hazards requires dynamic simulations under an array of multiple inputs from different sources. Uncertainty is associated with meteorological forecast and calibration of the model parameters. Software uncertainty also derives from the data transformation models (D-TM) needed for predicting hazard behaviour and its consequences. On the other hand, social contributions have recently been recognized as valuable in raw-data collection and mapping efforts traditionally dominated by professional organizations.
Here an architecture overview is proposed for adaptive and robust modelling of natural hazards, following the \textit{Semantic Array Programming} paradigm to also include the distributed array of social contributors called \textit{Citizen Sensor} in a semantically-enhanced strategy for D-TM modelling. The modelling architecture proposes a multi-criteria approach for assessing the array of potential impacts with qualitative rapid assessment methods {based on a \textit{Partial Open Loop Feedback Control} (POLFC) schema and complementing more traditional and} accurate a-posteriori assessment. We discuss the computational aspect of environmental risk modelling using {array-based} parallel paradigms on \textit{High Performance Computing} (HPC) platforms, in order for the implications of urgency to be introduced into the systems (Urgent-HPC).
\bigskip

\end{abstract}

\subsubsection*{1. INTRODUCTION: CONTEXT, PITFALLS AND THE SCIENCE-POLICY INTERFACE}
\medskip

\noindent Europe experienced a series of particularly severe disasters in the {recent years \cite{Sippel_Otto_2014,Cirella_etal_2014}, with worrying potential impacts of similar disasters under future projected scenarios of economy, society and climate change \cite{Ciscar_etal_2013,Ciscar_etal_2014}. They range} from flash floods {\cite{Dankers_Feyen_2008,Gaume_etal_2009,Marchi_etal_2010}} and severe storms in Western Europe {with an expected increasing intensity trend \cite{Feser_etal_2014}}, large-scale floods in Central Europe{ \cite{Jongman_etal_2014},} volcanic ash cloud{s} \cite{Self_2006,Swindles_2011,Gramling_2014} {(e.g. }after the Eyjafjallajökull eruption{)}, large forest fires in Portugal and Mediterranean countries {\cite{Allard_etal_2013,Schmuck_etal_2014}. Biological invasions such as emerging plant pests and diseases have the potential to further interact e.g. with wildfires \cite{Nijhuis_2012} and to impact on ecosystem services \cite{Boyd_etal_2013} and economy with substantial uncertainties~\cite{Venette_etal_2012}}.

It should be underlined that these recent highlights are set in the context of systemic changes in key sectors \cite{Maes_etal_2013,EC_2013,EC_2013b} which overall may be expected to at least persist in the next decades. As a general trend, demands on the EU's resilience in preparedness and disaster response capacity are likely to increase, as the impacts of disasters continue to grow both in size and frequency, even considering only the growing exposure (societal factors) \cite{Barredo_2007,Barredo_2010}.
The aforementioned examples of disturbances are often characterised by non-local system feedbacks and off-site impacts which may connect multiple natural resources (system of systems) \cite{Evans_etal_2012,Phillis_Kouikoglou,Langmann_2014}. In this particular multifaceted context \cite{Gottret_2001,Hagmann_2001,Zhang_2004}, landscape \cite{Estreguil_etal_2013} and ecosystem dynamics show intense interactions with disturbances \cite{Turner2010}. 

As a consequence, classical disciplinary and domain-specific approaches which might be perfectly suitable at local-scale may easily result in unacceptable simplifications within a broader context. A broad perspective is also vital for investigating future natural-hazard patterns at regional/continental scale and adapting preparedness planning \cite{Van_Westen_2013,Urban_etal_2012,Baklanov_2007}. The complexity and uncertainty associated with these interactions -- along with the severity and variety of the involved impacts \cite{Steffen_2011} -- urge robust, holistic coordinated \cite{White_etal_2012} and transparent approaches \cite{deRigo2013,deRigoSubm}. At the same time, the very complexity itself of the control-system problems involved  \cite{Lempert_2002,Rammel_2007,van_der_Sluijs_2012} may force the analysis to enter into the region of deep-uncertainty \cite{de_Rigo_etal_IFIP2013}. 

The mathematization of systems in this context as a formal control problem should be able to establish an effective science-policy interface, which is \textit{not} a trivial aspect. This is easily recognised even just considering the peculiarities -- which have been well known for a long time -- of geospatially-aware environmental data \cite{Guariso} and decision support systems \cite{GuarisoEDSS,Guariso_1985,Soncini_Sessa_2007}, their entanglement with growingly complex ICT aspects \cite{Guariso_1994,Casagrandi} and their not infrequent cross-sectoral characterisation. 
Several pitfalls may degrade the real-world usefulness of the mathematization/implementation process. While it is relatively intuitive how a poor mathematization with a too simplistic approach might result in a failure, subtle pitfalls may lie even where an ``appropriately advanced'' theoretical approach is proposed.  
Mathematization should resist \textit{silo thinking} \cite{Cole_2010,Sterman_2002} temptations such as academic solution-driven pressures \cite{deRigoSubm,Weichselgartner_2010} to force the problem into fashionable ``hot topics'' of control theory: robust approximations of the real-world broad complexity may serve egregiously instead of state-of-art solutions of oversimplified problems. 

Other long-lasting academic claims are ``towards'' fully automated scientific workflows in computational science, maybe including self-healing and self-adapting capabilities of the computational models implementing the mathematization. These kinds of claims might easily prompt some irony \cite{Bainbridge_1983} among experienced practitioners in wide-scale transdisciplinary modelling for environment (WSTMe, \cite{GeoSemAP_2013}) as a never-ending research Pandora's box with doubtful net advantages \cite{Stensson_2013}. Complex, highly uncertain and sensitive problems for policy and society, as WSTMe problems typically are, will possibly never be suitable for full automation: even in this family of problems, ``humans will always be part of the computational process'' \cite{Russell_2003} also for vital accountability aspects \cite{Anderson_2003}. 

While a certain level of autonomic computing \cite{Kephart_2003} capabilities might be essential for the evolvability and robustness of WSTMe (in particular, perhaps, a higher level of semantic awareness in computational models and a self-adapting ability to scale up to the multiple dimensions of the arrays of data/parameters; see next section), here the potential pitfall is the illusion of \textit{fully automating} WSTMe. The domain of applicability of this puristic academic \textit{silo} -- although promising for relatively simple, well-defined (and not too policy-sensitive) case studies -- might be intrinsically too narrow for climbing up to deal with the \textit{wicked problems} typical of complex environmental systems \cite{van_der_Sluijs_2005,Frame_2008,McGuire_2010}. 

The discussed pitfalls might deserve a brief summary. First, perhaps, is the risk of ``solving the wrong problem precisely'' \cite{Bea_etal_2009} by neglecting key sources of uncertainty -- e.g. unsuitable to be modelled within the ``warmly supported'' solution of a given research group. During emergency operations, the risks of providing a ``myopic decision support'' should be emphasised; i.e. suggesting inappropriate actions \cite{Adams_Hester_2012} -- e.g. inaction or missing precaution -- due to the potential overwhelming lack of information \cite{Larsson_etal_2010} or the oversimplification/underestimation of potential chains of impacts due to the lack of computational resources for a decent (perhaps even qualitative and approximate) rapid assessment of them. 

Overcoming these pitfalls is still an open issue. Here, we would like to contribute to the debate by proposing the integrated use of some mitigation approaches. We focus on some general aspects of the \textit{modelling architecture} for the computational science support, in order for emergency-operators, decision-makers, stakeholders and citizens to be involved in a participatory \cite{Innocenti_Albrito_2011} information and decision support system which assimilates uncertainty and precaution \cite{Ravetz_2004,van_der_Sluijs_2005}. Since no silver bullet seems to be available for mitigating the intrinsic wide-extent of complexity and uncertainty in environmental risk modelling, an array of approaches is integrated and the computational aspects are explicitly connected with the supervision and distributed interaction of human expertise. This follows the idea that the boundary between classical control-theory management strategies for natural resources and hazards (driven by automatic control problem formulations -- ``minimize the risk score function'') and scenario modelling under deep-uncertainty (by e.g. merely supporting emergency-operators, decision-makers and risk-assessors with understandable information -- ``sorry, no such thing as a risk score function can be precisely defined'') is fuzzy. Both modelling and management aspects may be computationally intensive and their integration is a transdisciplinary problem (\textit{integrated natural resources modelling and management}, INRMM \cite{INRMM}).

\medskip

\subsubsection*{2. ENVIRONMENTAL RISK MODELLING - ARCHITECTURE}

\smallskip
\noindent Figure 1 illustrates a general modelling conceptualization {where} the interactions among natural hazard behaviour, related transdisciplinary impacts, risk management and control strategies are taken into account. The special focus on the many sources of uncertainty \cite{deRigo2013} leads to a robust semantically-enhanced modelling architecture based on the paradigm of Semantic Array Programming (SemAP) \cite{deRigo2012,deRigo2012b}, with an emphasis on the array of input, intermediate and output data/parameters and the array of data-transformation modules (D-TM) dealing with them. 

Arrays of hazard models $h_j^{\zeta_f}(\cdot)$, dynamic information forecasts $X^{\,\zeta_X}$ (i.e. meteorology) and static parametrisation $\theta^{\,\zeta_\theta}$ (i.e. spatial distribution of land cover) are considered. Their multiplicity derives from the many sources on uncertainty $\zeta = \{\zeta_f, \zeta_X, \zeta_\theta\}$ which affect their estimation (or implementation, for the D-TM software modules $f_i^{\zeta_f}(\cdot)$ which are the building blocks of the hazard models $h_j^{\zeta_f}(\cdot)$). 

Furthermore, during emergency modelling support the lack of timely and accurate monitoring systems over large spatial extents (e.g. at the continental scale) may imply a noticeable level of uncertainty to affect possibly even the location of natural hazards (\textit{geoparsing}
 \cite{Corti_etal_2012} uncertainty). This peculiar information gap may be mitigated by integrating remote sensing (e.g. satellite imagery) with a distributed array of social contributors (Citizen Sensor \cite{Sheth_2009,Zhang_etal_2011,Adam_etal_2012}), exploiting mobile applications (Apps) and online social networks \cite{Fraternali_2012}. Remote sensing and the Citizen Sensor are here designed to cooperate by complementing accurate (but often less timely) geospatial information with distributed alert notifications from citizens, which might be timely but not necessarily accurate. Their safe integration implies the supervision of human expertise, even if the task may be supported by automatic tools \cite{Aseretto_submitted}. 
 Assessing the evolution in the timespan $\mathcal{U}^{\,t} = [t_\text{begin},t_\text{end}]$ of a certain hazard event for the associated array of impacts $C^{k,t}$ may be also complex (e.g. \cite{de_Rigo_etal_IFIP2013,Bosco_etal_2013,DiLeo_etal_2013}). In particular, the array of impacts is often irreducible to a unidimensional quantity (e.g. monetary cost) \cite{Ackerman_Heinzerling_2002,Gasparatos_2010}. 

\begin{center}
\includegraphics[width=\textwidth]{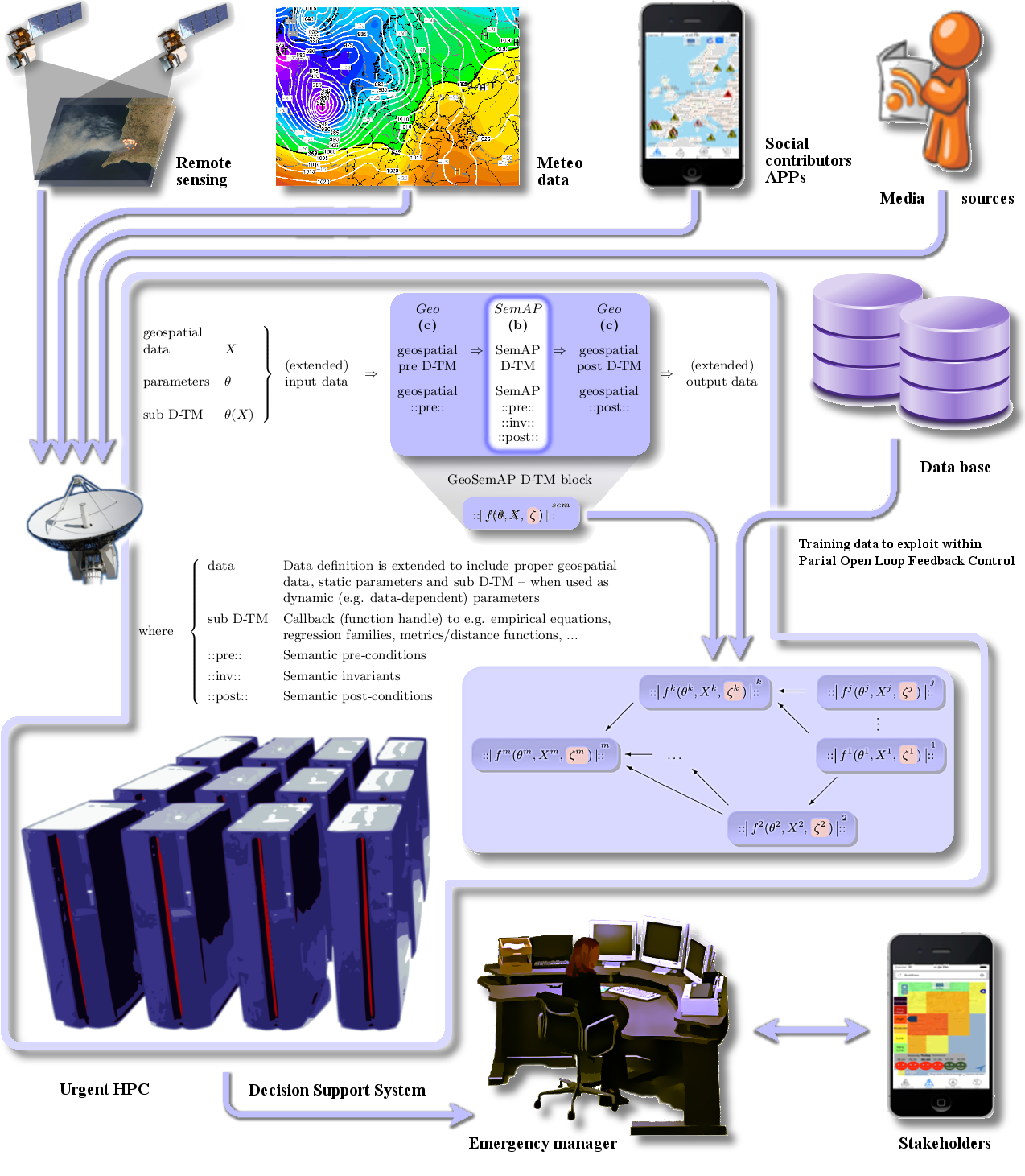}

{\small \textbf{Figure 1} - Modular architecture for environmental risk modelling. Based on Urgent~HPC, it follows the Semantic Array Programming paradigm (image adapted from \cite{deRigo2013,GeoSemAP_2013}) integrating as inputs remote sensing, meteo data and the Citizen Sensor.}

\bigskip
\end{center}
 
\noindent The analysis of non-trivial systems subject to environmental risk and natural resources management may naturally lead to multi-objective (multi criteria) control problems, which might benefit from advanced machine learning techniques for mitigating the involved huge computational costs \cite{de_Rigo_2001}. Indeed, the multiplicity of modelling dimensions (states; controls; uncertainty-driven arrays of parameters and scenarios; arrays of D-TM modules to account for software uncertainty) may easily lead to an exponential increase of the required computational processes (the so called ``curse of dimensionality''). A viable mitigation strategy might be offered by HPC tools (such as Urgent HPC \cite{RodriguezAseretto_etal_2009,Cencerrado_etal_2009,Yoshimoto2012}) in order to sample {high-dimensional modelling space with a proper method}.

\smallskip
\begin{mdframed}[roundcorner=10pt,backgroundcolor=Boxbackground]
\newcommand{\mydot}{\ensuremath{\bullet\;}}%
\newcommand{\mydotTwo}{\ensuremath{\qquad\triangleright\;}}
{\vspace{1mm}\begin{center}
{\bf Box 1} -- In a nutshell.
\end{center}\vspace{-2mm}

\footnotesize 
\begin{description}[leftmargin=0.8cm, style=sameline]
\item[Context] Demands on the EU's resilience in preparedness and disaster response capacity are likely to increase, as the impacts of disasters continue to grow.

\mydot{} Classical disciplinary and domain-specific approaches which might be perfectly suitable at local-scale may result in unacceptable simplifications in a broader context.

\item[Pitfalls] Mathematization of systems in this contex as a formal control problem should be able to establish an \textit{effective} science-policy interface. Academic \textit{silo thinking} should stop advertising solution-driven oversimplification to fit control theory ``hot topics''.

\mydot{} Although in this family of problems ``humans will always be part of the computational process'' (despite any academic potential illusion of fashionable \textit{full automation}),

\mydot{} evolvability (for adapting models to new emerging needs and knowledge) and robustness (for supporting uncertainty-aware decision processes) would still need 

\mydotTwo{} a higher level of semantic awareness in computational models and 

\mydotTwo{} a self-adapting ability to scale up to the multiple dimensions of the arrays of \\ \phantom{\mydotTwo{}} data/parameters.

\item[Multiplicity: uncertainty and complexity] In this context, the boundary between classical control-theory management strategies for natural resources and hazards and scenario modelling under deep-uncertainty is fuzzy (INRMM). 

\mydot{} A key aspect of soundness relies on explicitly considering the multiple dimensions of the problem and the array of uncertainties involved.

\mydot{} As no silver bullet seems to be available for reliably attacking this amount of uncertainty and complexity, an integration of methods is proposed.

\item[Mitigating with an integrated approach] Array programming is well-suited for easily managing a multiplicity of arrays of hazard models, dynamic input information, static parametrisation and the distribute array of social contributions (Citizen~Sensor).

\mydot{} \textit{Array-based abstract} -- thus better scalable -- \textit{modularisation} of the data-trans\-formations (D-TM), and a \textit{semantically-enhanced} design of the D-TM structure and interactions (Semantic Array Programming) is proposed to consider also the array of uncertainties (data, modelling, geoparsing, software uncertainty) and the array of criteria to assess the potential impacts associated with the hazard scenarios.

\mydot{} The unevenly available information during an emergency event may be
efficiently exploited by means of a POLFC schema.

\mydot{} Its demanding computations may become affordable during an emergency event with an appropriate array-based parallelisation strategy within Urgent-HPC.
\vspace{1.5mm}
\end{description}

}
\end{mdframed}

\noindent {SemAP can simplify WSTMe modelling of nontrivial static \cite{de_Rigo_Bosco_2011,de_Rigo_etal_EZ2013} and dynamic \cite{de_Rigo_etal_IFIP2013,DiLeo_etal_2013,RodriguezAseretto_etal_2013} geospatial quantities. Under the SemAP paradigm, the generic $i$-th D-TM module $Y_i = f_i(\theta_i,X_i)$} is subject to the semantic checks $\text{\textit{sem}}_i$ as pre-, post-conditions and invariants on {the} inputs {$\theta_i,X_i$}, outputs {$Y_i$} and the D-TM itself {$f_i(\cdot)$}. 
The control problem is associated with the unevenly available dynamic updates of field measurements and other data related to an on-going hazard emergency. An \textit{Emergency Manager} may thus be interested in assessing the best control strategy $u^t(\cdot)$ given a set of impacts and their associated costs as they can be approximately estimated (rapid assessment) with the currently available data. 
This data-driven approach can be implemented as Partial Open Loop Feedback Control (POLFC) approach~\cite{Castelletti2008} for minimizing the overall costs associated with the natural hazard event, from the time $t \in \mathcal{U}^{\,t}$ onwards:

\vspace{-2.5mm}
\begin{equation}
\label{eq:multicriteria_policy}
u^t(\cdot) = {\operatorname{arg\,min}}_{u\, \in \,{ U }^{ \, u }_{ t\, ,\, t_{ { end } } }} \left[ \mathcal{C}^{\,1,\,t} \mathcal{C}^{\,2,\,t} \cdots \mathcal{C}^{\,k,\,t} \cdots \mathcal{C}^{\,n,\,t} \right]
\end{equation}

\noindent where {the $k$-th} cost $\mathcal{C}^{\,k,\,t}$ is linked to the corresponding impact assessment criterion. {This POLFC schema within the SemAP paradigm may be considered a semantically-enhanced dynamic data-driven application system (DDDAS) \cite{de_Rigo_etal_IFIP2013,DiLeo_etal_2013,RodriguezAseretto_etal_2013}.} Finally, the Emergency Manager may communicate the updated scenarios of the emergency evolution (by means of geospatial maps and other executive summary information) in order for decision-makers and stakeholders to be able to assess the updated multi-criteria pattern of costs and the preferred control options. 
This critical communication constitutes the science-policy interface and must be as supportive as possible. It is designed to exploit web map services (WMS) \cite{McInerney2012,Bastin2012} (on top of the underpinning free software for WSTMe, e.g. \cite{RodriguezAseretto_EFDAC2013}) which may be accessed in a normal browser or with specific Apps for smart-phones~\cite{Aseretto_submitted}.

\medskip
\subsubsection*{3. CONCLU{DING} REMARKS}
\smallskip

\noindent NSF Cyberinfrastructure Council report reads: \textit{While hardware performance has been growing exponentially - with gate density doubling every 18 months, storage capacity every 12 months, and network capability every 9 months - it has become clear that increasingly capable hardware is not the only requirement for computation-enabled discovery. Sophisticated software, visualization tools, middleware and scientific applications created and used by interdisciplinary teams are critical to turning flops, bytes and bits into scientific breakthroughs} \cite{NSFC2007}. Transdisciplinary environmental problems such as the ones dealing with complexity and deep-uncertainty in supporting natural-hazard emergency might appear as seemingly intractable \cite{Altay_Green_2006}. Nevertheless, approximate rapid-assessment based on computationally intensive modelling may offer a new perspective at least able to support emergency operations and decision-making with qualitative or semi-quantitative scenarios. Even a partial approximate but timely investigation on the potential interactions of the many sources of uncertainty might help emergency managers and decision-makers to base control strategies on the best available -- although typically incomplete -- sound scientific information. In this context, a key aspect of soundness relies on explicitly considering the multiple dimensions of the problem and the array of uncertainties involved. As no silver bullet seems to be available for reliably attacking this amount of uncertainty and complexity, an integration of methods is proposed, inspired by their promising synergy. Array programming is perfectly suited for easily managing a multiplicity of arrays of hazard models, dynamic input information, static parametrisation and the distribute array of social contributions (Citizen Sensor). The transdisciplinary nature of complex natural hazards -- their need for an unpredictably broad and multifaceted readiness to robust scalability -- may benefit (1) from a disciplined \textit{abstract modularisation} of the data-transformations which compose the models (D-TM), and (2) from a \textit{semantically-enhanced} design of the D-TM structure and interactions. These two aspects define the Semantic Array Programming (SemAP, \cite{deRigo2012,deRigo2012b}) paradigm whose application -- extended to geospatial aspects \cite{GeoSemAP_2013} -- is proposed to consider also the array of uncertainties (data, modelling, geoparsing, software uncertainty) and the array of criteria to assess the potential impacts associated with the hazard scenarios. The unevenly available information during an emergency event may be efficiently exploited by means of a partial open loop feedback control (POLFC, \cite{Castelletti2008}) schema, already successfully tested in this integrated approach \cite{de_Rigo_etal_IFIP2013,DiLeo_etal_2013,RodriguezAseretto_etal_2013} as a promising evolution of adaptive data-driven strategies \cite{RodriguezAseretto_etal_2008}. Its demanding computations may become affordable during an emergency event with an appropriate array-based parallelisation strategy within Urgent-HPC.

\begin{spacing}{0.85}

\begin{nohyphens}

\end{nohyphens}

\end{spacing}

\end{document}